\documentclass[prl,amsfonts,amssymb,floats,twocolumn,superscriptaddress,showpacs,aps]{revtex4}

\usepackage[final,dvips]{epsfig}

\begin{document}

\title[]
{Quantum Phase Transitions in the Bosonic 
Single-Impurity Anderson Model
}

\author{Hyun-Jung Lee}
%\affiliation{\mbox{Theoretische Physik III, Elektronische Korrelationen und
%Magnetismus, Universit\"at Augsburg,
%86135 Augsburg, Germany}}
\author{Ralf Bulla}
\affiliation{\mbox{Theoretische Physik III, Elektronische Korrelationen und
Magnetismus, Universit\"at Augsburg,
86135 Augsburg, Germany}}

\date{\today}

\begin{abstract}
We consider a quantum impurity model in which a bosonic impurity
level is coupled to a non-interacting bosonic bath, with the bosons
at the impurity site subject to a local Coulomb repulsion $U$. Numerical
renormalization group calculations for this bosonic single-impurity
Anderson model reveal a zero-temperature phase diagram where
Mott phases with reduced charge fluctuations are separated from a
Bose-Einstein condensed phase by lines of quantum critical
points. We discuss possible realizations of this model, such as
atomic quantum dots in optical lattices. Furthermore, the bosonic
single-impurity Anderson model appears as an effective impurity model
in a dynamical mean-field theory of the Bose-Hubbard model.
\end{abstract}

\maketitle

The focus of this work is the physics of a bosonic impurity state
coupled to a non-interacting bosonic environment modeled
by the Hamiltonian
\begin{eqnarray}
   H &=& \varepsilon_0 b^\dagger b  
       + \frac{1}{2} U b^\dagger b \left( b^\dagger b -1\right)
    \nonumber \\
     & & + \sum_k \varepsilon_k b^\dagger_k b_k 
       + \sum_k V_k \left( b^\dagger_k b +  b^\dagger b_k\right) \ .
\label{eq:bsiam}
\end{eqnarray}
The energy of the impurity level (with operators $b^{(\dagger)}$)
is given by $\varepsilon_0$; the parameter $U$ is the local Coulomb
repulsion acting on the bosons at the impurity site. The impurity
couples to a bosonic bath via the hybridization $V_k$, with the
bath degrees of freedom given by the operators $b_k^{(\dagger)}$
with energy $\varepsilon_k$.

We term the system defined by eq.~(\ref{eq:bsiam})
the `bosonic single-impurity Anderson model' (bosonic siAm), in analogy to the
standard (fermionic) siAm \cite{Hewson}
which has a very similar structure except that all fermionic operators
are replaced by bosonic ones. Furthermore, we do not consider internal
degrees of freedom of the bosons, such as the spin (an essential
ingredient in the fermionic siAm).

There are various ways to motivate the study of the model 
eq.~(\ref{eq:bsiam}). From a purely theoretical point of view it is
interesting to compare the physics of the fermionic and bosonic
versions of the siAm. Of course, there are striking differences
between these two models: there is no direct bosonic analog
of local moment formation and screening of these local
moments at low temperatures, characteristics of the
fermionic Kondo effect \cite{Fritz:2005}.
On the other hand, the bosonic model allows for a Bose-Einstein
condensation (BEC), at least in a certain parameter space (see
Fig.~\ref{fig:phd} below), 
a phenomenon which is clearly absent in the fermionic
model.

There are certain similarities
between the fermionic and the bosonic model concerning the role
of the local Coulomb repulsion $U$: increasing the value of $U$ can
induce a quantum phase transition from a phase with screened spin
to a local moment phase in the fermionic case \cite{soft-gap}, 
while it can induce
a quantum phase transition from a BEC phase to a `Mott phase' in the
bosonic case, as shown below.

Another motivation for studying the bosonic siAm comes from a
treatment of the Bose-Hubbard model within dynamical mean-field
theory (DMFT) \cite{dmft}. Although such an investigation has not been pursued 
so far, it is clear that the effective impurity model onto
which the Bose-Hubbard model is mapped will have a similar
form as eq.~(\ref{eq:bsiam}) 
(see also Ref.~\onlinecite{Bulla:2006}). 
An obvious application of such a DMFT
treatment would then address the Mott transition in
the Bose-Hubbard model \cite{bhm} which must have its counterpart on the
level of the effective impurity model -- and the `Mott transition'
on the impurity level is precisely what we are looking at here.

Finally, a physical system described by the bosonic siAm could
be directly realized in optical lattices, where the laser fields are 
tuned in such a way that a single `impurity' site is formed within
an approximately unperturbed lattice system (`atomic quantum
dots', see Refs.~\onlinecite{Recati:2005,Jaksch:2005}). Provided the 
Coulomb repulsion at all sites except the impurity can be neglected,
the corresponding model can be directly mapped onto the model
eq.~(\ref{eq:bsiam}). Present theoretical studies of atomic 
quantum dots focus, however, on a coupling between impurity and 
excitations of the superfluid environment \cite{Recati:2005},
for which a description in terms of the spin-boson model is more
appropriate.

For the calculations presented in this paper we use 
the numerical renormalization group (NRG) originally developed by Wilson
for the Kondo problem \cite{Wil75}. This method
has been shown to give very accurate
results for a broad range of impurity models,  including
the fermionic siAm \cite{Kri80,Hewson} but also quantum impurities
with coupling to a bosonic bath \cite{BTV,BLTV}. Here we employ the
bosonic extension of the NRG (bosonic NRG) to study the model
eq.~(\ref{eq:bsiam}). We shall present ground state phase diagrams
of the bosonic siAm, identify fixed points and discuss quantum
phase transitions between the fixed points. Although far from
comprehensive, our results indicate that the bosonic siAm
shows a variety of interesting properties which deserve to be studied
in greater detail in the future.

Before we come to the results from the bosonic NRG, let us discuss
some general properties and trivial limits of the model
eq.~(\ref{eq:bsiam}).

Similar to other quantum impurity models, the influence of the bath
on the impurity is completely specified by the bath spectral
function
\begin{equation}
   \Delta(\omega) = \pi \sum_k V_k^2 \delta(\omega-\varepsilon_k) \ .
\end{equation}
Here we assume that $\Delta(\omega)$ can be parametrized by
a powerlaw for frequencies up to a cutoff $\omega_{\rm c}$ (we
set $\omega_{\rm c}=1$ in the calculations)
\begin{equation}
   \Delta(\omega) =  2\pi\, \alpha\, \omega_{\rm c}^{1-s} 
   \, \omega^s\,,~ 0<\omega<\omega_c\, \ .
\label{eq:power}
\end{equation}
The parameter $\alpha$ is the dimensionless coupling constant
for the impurity-bath interaction.
This form of $\Delta(\omega)$ is certainly not the most 
general one and specific applications
(within a bosonic DMFT or for an impurity level in an optical lattice)
will lead to additional structures in $\Delta(\omega)$.

The NRG calculations can be both performed for a canonical and
a grandcanonical ensemble. Here we present only results for
a grandcanonical ensemble for which we set the chemical potential
of the bath to $\mu = 0$. This means that a Bose-Einstein condensation
of the whole system (impurity plus bath) can only be induced by the
coupling to the impurity. This can already be understood from the
non-interacting case, $U=0$: here a direct diagonalization of the
bosonic siAm shows that with increasing $\alpha$, a localized state
with {\em negative} energy separates out of the continuum at a critical
value $\alpha=\alpha_{\rm c}$. The BEC then occurs via populating this
localized state, a feature which can also be observed in the numerical
calculations.

The other trivial limit of the bosonic siAm is the decoupled impurity,
$\alpha=0$. 
The succession of quantum phase transitions as shown in Fig.~\ref{fig:phd}
for $\alpha=0$
can be easily understood from the dependence of the many-particle
levels on the parameters $\varepsilon_0$ and $U$.  The transition
occurs for $\varepsilon_0/U=-n_{\rm imp}$ when the energies of the states 
with $n_{\rm imp}$ and 
$n_{\rm imp}+1$ electrons are degenerate.

The full phase diagram Fig.~\ref{fig:phd} 
is calculated with the bosonic NRG \cite{BLTV}.
In this approach, the frequency range of the bath spectral function
$[0,\omega_{\rm c}]$ is divided into intervals 
$[\omega_{\rm c}\Lambda^{-(n+1)},\omega_{\rm c}\Lambda^{-n}]$,
$n=0,1,2,\ldots$, with $\Lambda$ the NRG discretization parameter
(we use $\Lambda=2.0$ for all the results shown in this paper).
The continuous spectral function within these intervals is
approximated by a single bosonic state and the resulting
discretized model is then mapped onto a semi-infinite chain
with the Hamiltonian
\begin{eqnarray}
   H &=& \varepsilon_0 b^\dagger b  
       + \frac{1}{2} U b^\dagger b \left( b^\dagger b -1\right)
       + V \left( b^\dagger \bar{b}_0 + \bar{b}_0^\dagger b \right)
    \nonumber \\
     & & + \sum_{n=0}^\infty \varepsilon_n \bar{b}^\dagger_n \bar{b}_n 
       + \sum_{n=0}^\infty t_n \left( 
             \bar{b}^\dagger_n \bar{b}_{n+1}  + 
             \bar{b}^\dagger_{n+1} \bar{b}_{n} \right) \ .
\label{eq:bsiam-chain}
\end{eqnarray}
Here the impurity couples to the first site of the chain via
the hybridization $V=\sqrt{2\alpha/(1+s)}$. The bath degrees of freedom are
in the form of a tight-binding chain with operators 
$\bar{b}^{(\dagger)}_n$, on-site energies $\varepsilon_n$,
and hopping matrix elements $t_n$ which both fall off
exponentially: $t_n,\varepsilon_n\propto \Lambda^{-n}$.

The chain model eq.~(\ref{eq:bsiam-chain}) 
is diagonalized iteratively starting with the
impurity site and adding one site of the chain in each iteration.
As for the application to the spin-boson model, only a finite
number $N_{\rm b}$ of basis states for the added site can be
taken into account, and, after diagonalizing the enlarged cluster,
only the lowest-lying $N_{\rm s}$ many-particle states are
kept for the subsequent iterations (here we use $N_{\rm b}=10-20$
and $N_{\rm s}=100-200$). The main technical difference to the
spin-boson model is that we can use the total particle-number
as a conserved quantity in the Hamiltonian eq.~(\ref{eq:bsiam}). 
Furthermore, the
renormalization group flow turns out to be considerably more stable
as in the calculations for the spin-boson model so that we can
easily perform up to $N=100$ iterations (a detailed account of the
technical details will appear elsewhere).

\begin{figure}[!t]
\epsfxsize=3.1in
\centerline{\epsffile{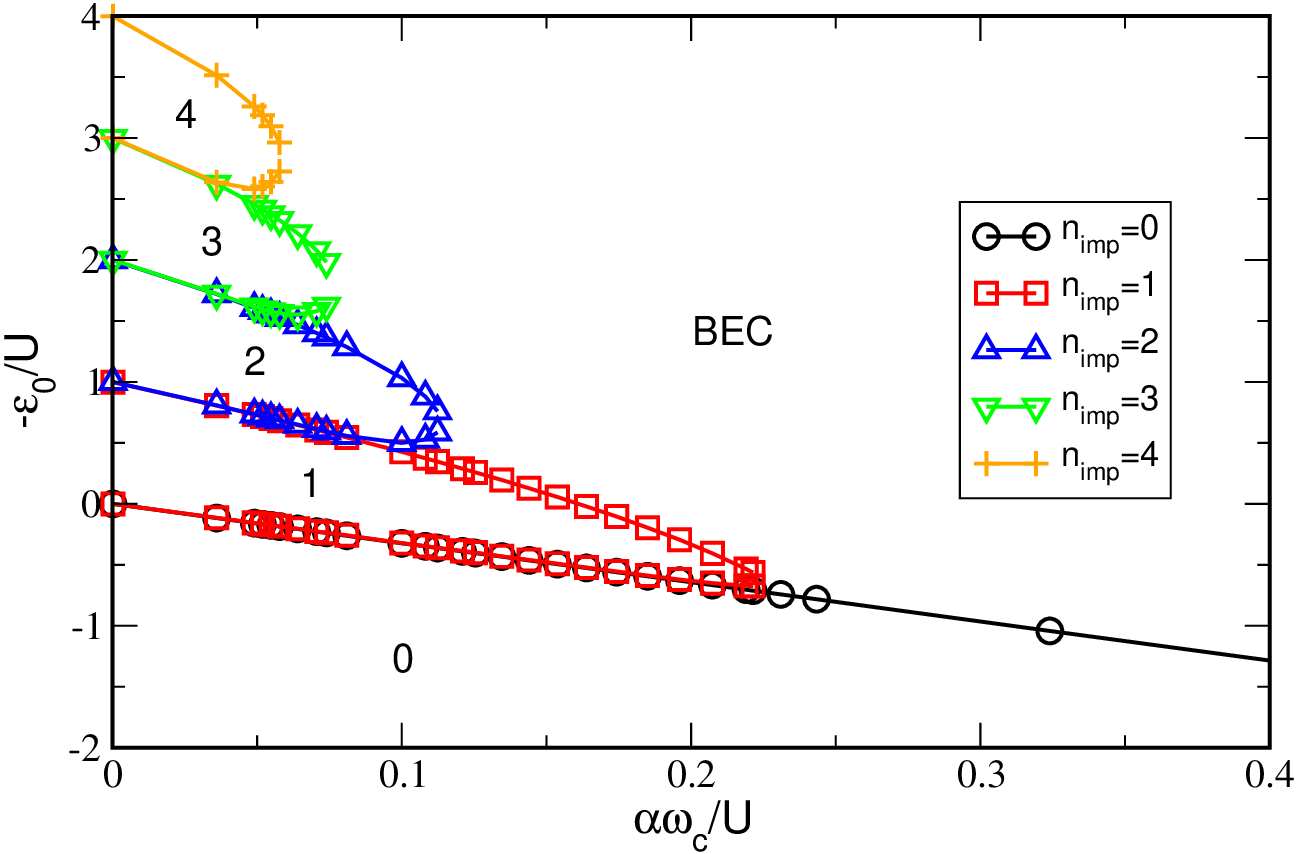}}
\caption{Zero-temperature phase diagram of the bosonic siAm
for bath exponent $s=0.6$ and fixed impurity Coulomb interaction
$U=0.5$. The different symbols denote the
phase boundaries between the Mott phases and the BEC phase. The Mott phases
are labeled by their occupation $n_{\rm imp}$ for $\alpha=0$. Only
the Mott phases with $n_{\rm imp}\le 4$ are shown. The
NRG parameters are $\Lambda=2.0$, $N_{\rm b}=10$,
and $N_{\rm s}=100$.
}
\label{fig:phd}
\vspace*{-0.3cm}

\end{figure}

Let us now discuss the $T=0$ phase diagram of the bosonic siAm,
Fig.~\ref{fig:phd}, calculated for fixed $U=0.5$ with the parameter
space spanned by the dimensionless
coupling constant $\alpha$ and the impurity energy $\varepsilon_0$.
We choose $s=0.6$ as the exponent of the powerlaw in $\Delta(\omega)$
(the $s$-dependence of the phase diagram is discussed in Fig.~\ref{fig:phd-s}
below). The phase diagram is characterized by a sequence of lobes
which we label by the impurity occupation $n_{\rm imp}$
at $\alpha=0$ where it takes integer values. We use the terminology
`Mott phases' for these lobes, due to the apparent similarity
to the phase diagram of the Bose-Hubbard model. The Mott phases
are separated from the BEC phase by lines of quantum critical points
which terminate (for $s=0.6$) at a finite value of  $\alpha$,
except for the $n_{\rm imp}=0$ phase where the boundary extends up to
infinite $\alpha$. These transition can be viewed as the impurity analogue
of the Mott transition in the lattice model, since it is the
local Coulomb repulsion which prevents the formation of the
BEC state.

\begin{figure}[!t]
\epsfxsize=3.1in
\centerline{\epsffile{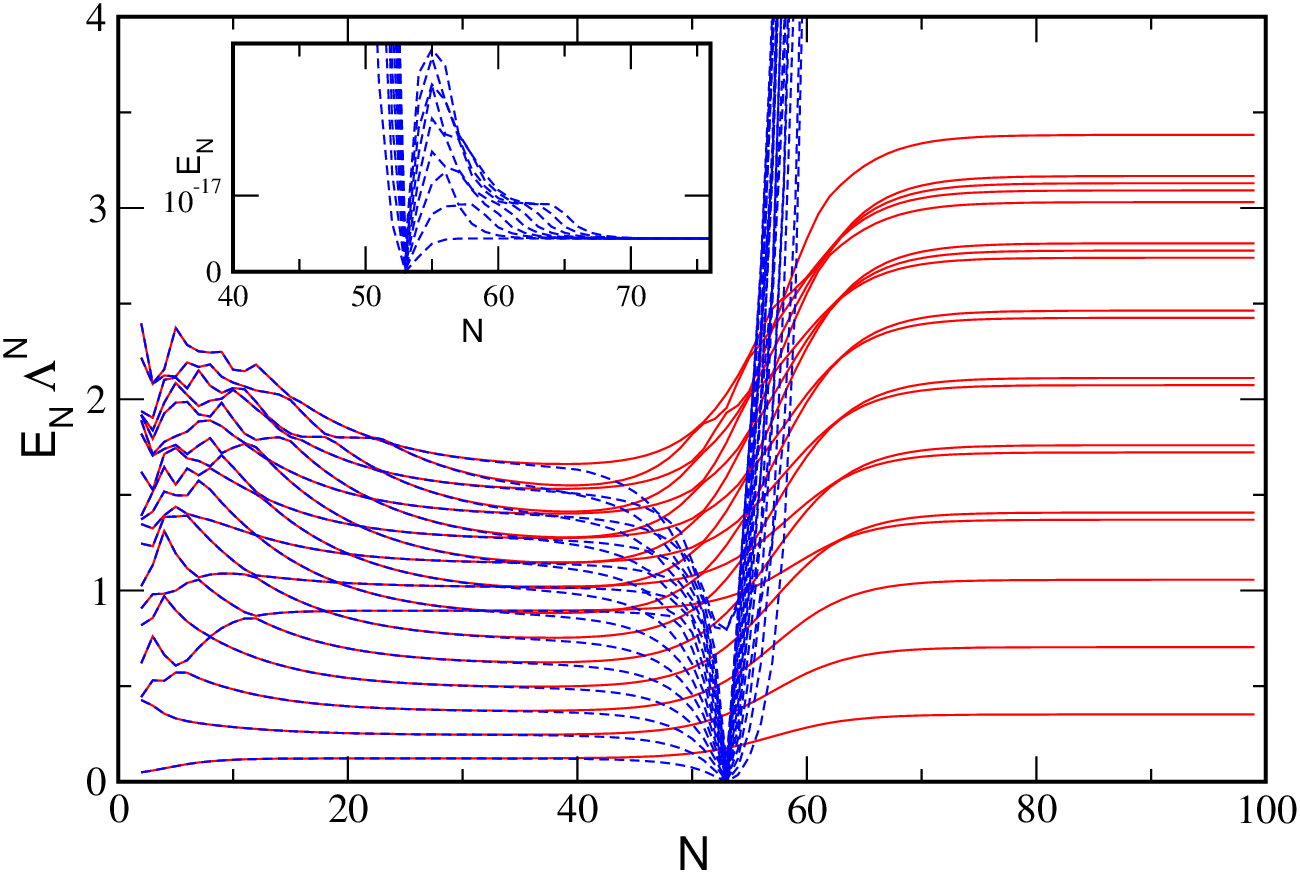}}
\caption{Flow diagram of the lowest lying many-particle levels $E_N$
versus iteration number $N$ for parameters $s=0.4$, $\alpha=0.007$,
$U=0.5$, and two values of $\varepsilon_0$ very close to the
quantum phase transition between the Mott phase with $n_{\rm imp}=2$
and the BEC phase. Both the quantum critical point and the Mott phase
appear as fixed points in this scheme whereas in the BEC phase, a
gap $\Delta_{\rm g}$ opens between the ground state and the first excited
state, see the inset where $E_N$ (instead of $E_N \Lambda^N$)
is plotted versus $N$.
}
\label{fig:flow}
\vspace*{-0.3cm}
\end{figure}

The phase diagram Fig.~\ref{fig:phd} is deduced from the
flow of the lowest-lying many-particle levels which allows
quite generally to identify fixed points and transitions
between these fixed points \cite{Kri80}. Figure \ref{fig:flow}
shows the flow for $s=0.4$, fixed $\alpha=0.007$ and $U=0.5$,
and two values of $\varepsilon_0$ very close to the quantum
phase transition. The solid lines belong to the Mott phase
with $n_{\rm imp}=2$ and display a crossover between two
fixed points: from an unstable quantum critical point
for iteration numbers $N\sim 20 - 40$ to a stable fixed point
for $N> 70$. Analysis of the stable fixed point shows that 
it can be described by a decoupled impurity with occupation
$n_{\rm imp}=2$ and a free bosonic bath given by the
decoupled chain. (The actual impurity occupation, however, differs
from the integer values, as discussed below.)
The structure of the quantum critical point,
on the other hand, is presently not clear but might be
accessible to perturbative methods as discussed in \cite{LBV}.

The dashed lines belong to the BEC phase where, apparently,
something dramatic is happening at $N\approx 55$. The divergence of 
the energies for all excited states seen in this plot is due
to the formation of a localized state, split off from the continuum
by an energy gap $\Delta_{\rm g}$,
very similar to the behavior for $U=0$. This energy gap takes
a finite value which is not renormalized
to zero as $\Lambda^{-N}$ as the levels of the other fixed points,
therefore the divergence of $\Delta_{\rm g}\cdot \Lambda^N$.
In the inset of Fig.~\ref{fig:flow} we plot $E_N$ 
(instead of $E_N \Lambda^N$) so that the development of the
gap $\Delta_{\rm g}\approx 4\times 10^{-18}$ appears as a 
plateau of the first excitated state. The extremely small value of the
gap is due to the tuning of the parameters very close to the
transition ($\Delta_{\rm g}$ vanishes at the transition).

Further analysis of the data shows that the ground state in the
BEC phase has an occupation number given by the maximum boson
number used in the iterative diagonalization. 

The fact that the unstable fixed point separating the Mott phase
from the BEC phase is indeed a quantum critical point can
be deduced from its non-trivial level structure, as mentioned
above, and the behavior of the crossover scale $T^*$.
Numerically we find that upon variation of $\varepsilon_0$
close to its critical value $\varepsilon_{0,\rm c}$, the
crossover scale vanishes with a powerlaw at the transition,
$T^*\propto |\varepsilon_0 - \varepsilon_{0,\rm c}|^\nu$
on both sides of the transition, with a non-trivial exponent
$\nu \approx 2.50 \pm 0.06$. This exponent is observed for all lines
of quantum critical points for fixed $s=0.4$ and preliminary results
show that $\nu=1/s$ for $0<s<1$.

\begin{figure}[!t]
\epsfxsize=3.0in
\centerline{\epsffile{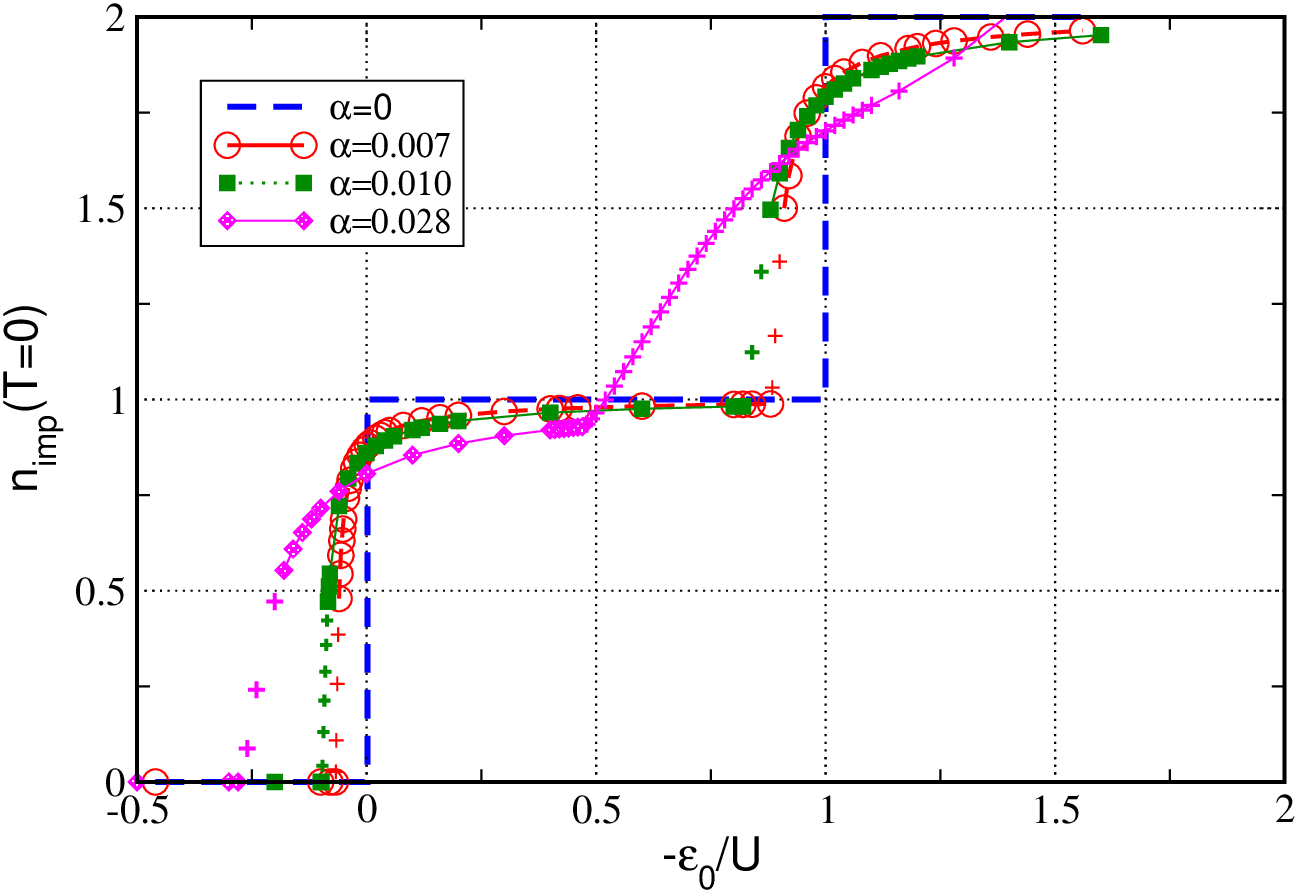}}
\caption{Impurity occupation $n_{\rm imp}$ as a function of
$\varepsilon_0$ for temperature $T=0$, $s=0.4$, $U=0.5$,
and various values of $\alpha$. The sharp steps for the
decoupled impurity $\alpha=0$ are rounded for any finite
$\alpha$. Symbols (crosses) correspond to
data points within the Mott (BEC) phases.
}
\label{fig:nimp}
\vspace*{-0.3cm}
\end{figure}

Let us now focus on the impurity occupation $n_{\rm imp}$ 
for temperature $T=0$. Figure \ref{fig:nimp} shows the dependence
of $n_{\rm imp}$ on $\varepsilon_0$ for  $s=0.4$, $U=0.5$,
and various values of $\alpha$. The symbols indicate that the
parameters lie within the Mott phase whereas the crosses are for
the BEC phase. Both sets of lines form a continuous curve
but it is unclear at the moment whether the data in the BEC phases
are reliable (due to the special features in the flow diagram discussed
above).

The terminology we use
here suggests an integer occupation throughout the Mott phase,
as for the Bose-Hubbard model, but for the bosonic siAm 
$n_{\rm imp}$ deviates from the integer values as soon as
the coupling to the bath is finite, see Fig.~\ref{fig:nimp}.
This is to be expected since for a gapless bath spectral function
$\Delta(\omega)$, the charge fluctuations on the impurity site
cannot be completely suppressed. Indeed we observe that for
increasing the value of the bath exponent $s$, the 
$n_{\rm imp}(\varepsilon_0)$-curve gets closer to the step function.
At this point one can speculate about the possible development
of $\Delta(\omega)$ in a DMFT treatment of the Mott phase. The
self-consistency might generate a bath spectral function
with a gap and the impurity occupation might then turn into
the step function expected for the lattice model.

The precise shape of the boundaries in the phase diagram
Fig.~\ref{fig:phd} depends on the form of $\Delta(\omega)$
for {\em all} frequencies. Here we stick to the powerlaw
form eq.~(\ref{eq:power}) and present the dependence of the
phase diagram on the bath exponent $s$ in Fig.~\ref{fig:phd-s}.
We observe that upon increasing the value of $s$, the areas 
occupied by the Mott phases extend to larger values of $\alpha$
and significantly change their shape. A qualitative change is
observed when the exponent approaches $s=1$. First of all, the
Mott phases appear to extend up to arbitrarily large values of 
$\alpha$. Furthermore, the BEC phase which separates the Mott phases
for $s<1$ and $\alpha>1$ is completely absent for $s=1$!
We do not yet have an explanation for this
observation and it would be interesting to find out whether the
absence of the BEC phase is due to the special form of
$\Delta(\omega)$, eq.~(\ref{eq:power}), or whether it is a generic
feature even when $\Delta(\omega)\propto\omega$ is only valid for
$\omega\to 0$.

The case $s=0$ (constant bath density of states) turns out to be difficult
to access numerically. An extrapolation of the phase boundaries for
values of $s$ in the range $0.1\ldots0.4$ to $s=0$ is inconclusive,
but the Mott phase is at least significantly suppressed in this limit.
%or: that a Mott phase still exists in this limit.

\begin{figure}[!t]
\epsfxsize=2.9in
\centerline{\epsffile{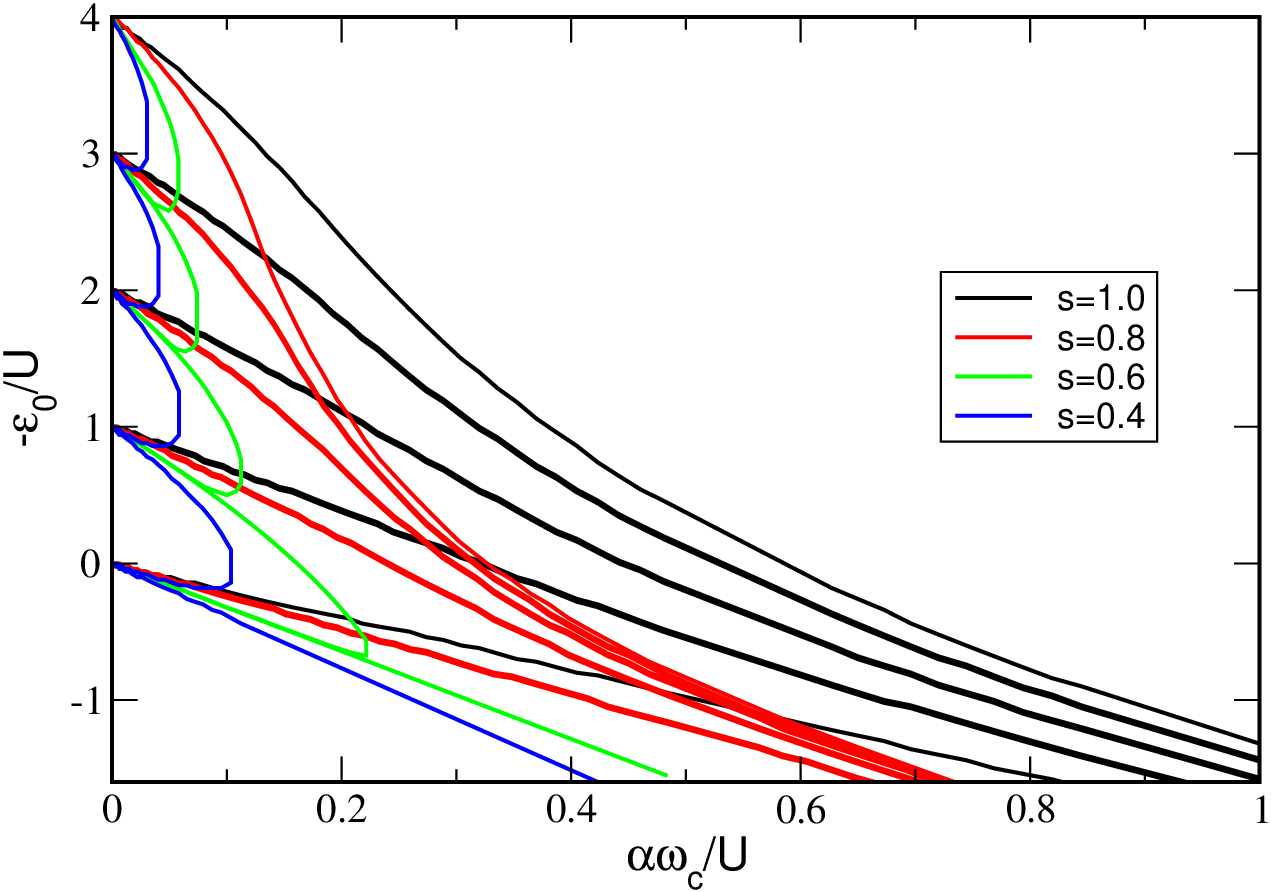}}
\caption{Zero-temperature phase diagram of the bosonic siAm
as in Fig.~\ref{fig:phd}, 
but now for different values of the bath exponent $s$.
For increasing value of $s$, the areas occupied by the Mott phases 
significantly change their shape and for $s=1$ it appears that each 
Mott phase extends up to arbitrarily large values of $\alpha$.
}
\label{fig:phd-s}

\vspace*{-0.3cm}
\end{figure}

To summarize, we have presented NRG calculations for the phase diagram 
and the impurity occupation of a bosonic version of the
single-impurity Anderson model. The phase diagram contains Mott
phases, in which the local Coulomb repulsion prevents Bose-Einstein
condensation, separated by the BEC phase by lines of quantum
critical points. The studies presented here are only the
starting point for a comprehensive investigation of the bosonic siAm
 and we are planning to calculate, for example,
physical properties at finite temperatures and dynamic
quantities (impurity spectral function and self-energy). The
latter will be of importance for a possible DMFT for the
Bose-Hubbard model, an approach which has not yet been
fully developed due to various conceptual problems. One issue is
the proper scaling of the Hamiltonian parameters in the limit
of infinite spatial dimensions \cite{DV}. For example, the model on a
hypercubic lattice as studied in Ref.~\onlinecite{FM} requires a scaling of
the hopping matrix elements as $1/d$ which leads to a static
mean-field theory. In addition, the bosonic DMFT in the
superfluid phase of the Bose-Hubbard model might generate a more
complex impurity model, the bosonic siAm introduced here would then
be applicable only within the Mott phases of the lattice model.

Finally, it would be interesting
to identify situations for atomic quantum dots in optical lattices
which can be described by the bosonic single-impurity Anderson model.

%%%%%%%%%%%%%%%%%%%%%%%%%%%%%%%%%%%%%%%%%%%%%%%%%%%%%%%%%%%%%%%%%%%%%%%%%

%

We would like to thank Krzysztof Byczuk, Jim Freericks, Matthias Vojta,
and Dieter Vollhardt
for helpful discussions.
This research was supported by the DFG through SFB 484.

%%%%%%%%%%%%%%%%%%%%%%%%%%%%%%%%%%%%%%%%%%%%%%%%%%%%%%%%%%%%%%%%%%%%%%%%%

\end{document}